\documentclass{svmult}

\usepackage{amssymb,amsmath}
\usepackage{graphicx}        
\usepackage{multicol}        
\usepackage[bottom]{footmisc}

\newcommand{\range}{\mathrm{Ran}\,}

\newcommand{\orbit}[1]{{\mathcal O}_{#1}}

\begin{document}

\title*{Open Subsystems of Conservative Systems}
\author{Alexander Figotin\inst{1}\and
Stephen P. Shipman\inst{2}}
\institute{University of Californina, Irvine, CA \ 92697 \texttt{afigotin@uci.edu}
\and Louisiana State University, Baton Rouge, LA \ 70803 \texttt{shipman@math.lsu.edu}}
%
%
\maketitle

\begin{abstract}
The subject under study is an open subsystem of a larger linear and conservative system and the way in which it is coupled to the rest of system.   Examples are a model of crystalline solid as a lattice of coupled oscillators with a finite piece constituting the subsystem, and an open system such as the Helmholtz resonator as a subsystem of a larger conservative oscillatory system. Taking the view of an observer accessing only the open subsystem we ask, in particular, what information about the entire system can be reconstructed having such limited access.  Based on the unique minimal conservative extension of an open subsystem, we construct a canonical decomposition of the conservative system describing, in particular, its parts coupled to and completely decoupled from the open subsystem. The coupled one together with the open system constitute the unique minimal conservative extension. Combining this with an analysis of the spectral multiplicity, we show, for the lattice model in particular, that \emph{only a very small part of all possible oscillatory motion of the entire crystal, described canonically by the minimal extension, is coupled to the finite subsystem.}  {\bfseries Keywords:}  open system, subsystem, conservative extension, coupling, delayed response, reconstructible.
\copyright A Figotin, SP Shipman
\end{abstract}

\section{Overview}\label{sectionOverview}

When one has to treat a complex evolutionary system involving a large number of, or
infinitely many, variables, it is common to reduce it to a smaller system by
eliminating certain ``hidden" variables.  The reduced system, involving only the ``observable" variables, becomes a non-conservative, or \emph{open system}, even if the underlying
system is conservative, or closed. This is not surprising since
generically any part of a conservative system interacts with the rest of it.
In the reduced system, the interaction with the hidden variables is encoded in its
dispersive dissipative (DD) properties. For
classical material media, including dielectric, elastic, and acoustic, the
interaction between proper fields and the matter, which constitutes the hidden part of the system, is encoded into the so-called
material relations, making them frequency dependent and consequently
making the open system dispersive and dissipative.

Often it is an open DD system, described by frequency-dependent material relations, that we are given to study, and the conservative system in which the open system is naturally embedded may be very complicated. A natural question is, how much information about the underlying conservative
system remains in the reduced open one?  The answer is
provided by the construction of the minimal conservative extension of the given DD system \cite{FigotinSchenker}, which is unique up to isomorphism.  This minimal extension is a part of the
entire conservative system---it is the part that is detectable by the open system through the coupling to the entire system.  We ask, how big a part of the original conservative system is this minimal extension?  This is a question we address in this paper.   The answer is clearly related to the nature of the coupling between the observable and hidden variables.   Although the term ``coupling" is commonly used to describe interactions, its precise meaning must be defined in each concrete problem. We make an effort to provide a general constructive mathematical framework for the treatment of the coupling.

In this paper, we concentrate on the detection of one part of a system by another, or, equivalently, the extent of reconstructibility of a conservative system from the dynamics of an open subsystem.  More detailed analysis of this problem as well as the study of the decomposition of open systems by means of their conservative extensions will be presented in another work.

\paragraph{Motivating Examples}


We have already mentioned the classical problems of electromagnetic, acoustic, and elastic waves in matter.   Detailed accounts of the construction of the minimal conservative extension are given in \cite{FigotinSchenker,Tip}.

Another important example is of an object coupled to a heat bath through surface contact.  It has been observed for crystalline solids that certain degrees of freedom do not contribute to the specific heat \cite[Section 3.1]{Gallavotti}, \cite[Section 6.4]{Huang}.  It appears that some of the admissible motions of the solid cannot be excited by the heat bath through the combination of surface contact and internal dynamics.  This can be explained though high multiplicity of eigenmodes arising from symmetries of the crystal.

A concrete toy model consists of an infinite three-dimensional lattice of point masses as the total system, each mass being coupled to its nearest neighbors by springs, and a finite cube thereof as the observable subsystem.  The coupling of the cube to the rest of the lattice takes place only between the masses on the surface of the cube and their nearest neighbors outside the cube.  We discuss this system in Example \ref{exampleLattice} below, in which we show that the cube is able to detect only a relatively small part of the entire lattice, the rest of which remains dynamically decoupled.

One more example is the phenomenon of anomalous acoustic or electromagnetic transmission through a material slab, or film, can also be viewed from the point of view of coupled systems.  The governing equation is the wave equation or the Maxwell system in space.  A leaky guided mode in a material slab interacts with plane wave sources from outside the slab, giving rise to anomalous scattering behavior \cite{Tikhodeev,ShipmanVenakides}.  A single mode of the slab constitutes a one-dimensional subsystem, which, under weak coupling to the ambient medium, say air, interacts with a portion of the entire system in space, decoupled from the rest.   We do not analyze this problem in this paper, but attempt to develop a framework for studying like problems.

\paragraph{List of Symbols} 

$H_1$, $H_2$, $\mathcal{H}$: Hilbert spaces\\
$v_1$, $v_2$, $\mathcal{V}$, $f_1$, $f_2$, $\mathcal{F}$: Hilbert space-valued functions of time\\
$\Gamma$, $\Omega_1$, $\Omega_2$, $\Omega$, $\mathring\Omega$, $\mathring\Gamma$: operators in Hilbert space\\
$a_1$, $a_2$: operator-valued functions of time\\
$\mathcal{O}$: orbit\\
$\range$: range\\
$\text{dim}$: dimension\\
$\mathbb{C}$: the complex number field\\
$\mathbb{Z}$: the ring of integers\\
$\mathcal{Q}$: a cube in $\mathbb{Z}^3$\\
$\Delta_j$: finite difference operators

\section{Open Systems Within Conservative Extensions}

Often an observable open system in a Hilbert space $H_1$ of the form
\begin{equation}\label{open}
\partial_t v_1(t) = -\I\Omega_1 v_1(t)
               - \int_0^\infty a_1(\tau)v_1(t-\tau)\,\D\tau
               + f_1(t) \quad \text{in  } H_1,
\end{equation}
in which $a_1(t)$ is the operator-valued delayed response, or retarded friction, function,
is known to be a subsystem of a linear conservative system in a larger Hilbert space $\mathcal H$, in which the dynamics are given by
\begin{equation}\label{dynamics1}
\partial_t \mathcal{V}(t) = -\I\Omega \mathcal{V}(t) + \mathcal{F}(t), \quad \mathcal{V}(t),\mathcal{F}(t)\in\mathcal{H},
\end{equation}
where $\Omega:\mathcal{H}\to\mathcal{H}$ is the self-adjoint frequency operator.  The structure of the open system within the conservative one can be seen by introducing the space $H_2$ of hidden variables, defined to be the orthogonal complement of $H_1$ in $\mathcal H$: $H_2 = \mathcal{H} \ominus H_1$.  With respect to the decomposition
$
\mathcal{H} = H_1\oplus H_2, 
$
$\Omega$ has the form 
\begin{equation}
\Omega = \left[  
\begin{array}{cc}
\Omega_1 & \Gamma \\ 
\Gamma^\dagger & \Omega_2%
\end{array}
\right]\, ,
\end{equation}
 in which $\Omega_1$ and $\Omega_2$ are the self-adjoint frequency operators for the internal dynamics in $H_1$ and $H_2$, and $\Gamma:H_2\to H_1$ is the coupling operator.   In this paper, we assume for simplicity that $\Gamma$ is bounded.  The results hold, essentially unchanged, for unbounded coupling; details of how to treat this case are handled in \cite{FigotinSchenker}.
The dynamics \eqref{dynamics1} with respect to the decomposition into observable and hidden variables become
\begin{align}
\partial _{t}v_1(t) & =-\I\Omega _{1}v_{1}\left(t\right) -\I\Gamma v_{2}\left( t\right) +f_{1}\left( t\right) , 
\quad v_1(t), f_1(t) \in H_1,
\label{smb} \\
\partial _{t}v_{2}\left( t\right) & =-\I\Gamma^{\dagger
}v_{1}\left( t\right) -\I\Omega _{2}v_{2}\left( t\right)
+f_{2}\left( t\right),
\quad v_2(t), f_2(t) \in H_2.
 \notag
\end{align}
Solving for $v(t)$ gives
\begin{equation}\label{reduced}
\partial_t v_1(t) = -\I\Omega_1 v_1(t)
               - \int_0^\infty \Gamma \E^{-\I\Omega_2 \tau} \Gamma^\dagger v_1(t-\tau)\,\D\tau
               + f_1(t) \quad \text{in  } H_1,
\end{equation}
from which we see that the delayed response function $a_1(t)$ is related to the dynamics of the hidden variables and the coupling operator by
\begin{equation}\label{delayedresponse}
a_1(t) = \Gamma \E^{-\I\Omega_2 t} \Gamma^\dagger,
\end{equation}
and it is straightforward to show that $a_1(t)$ satisfies the no-gain dissipation condition%
\begin{equation}\label{nogain}
\mathrm{Re}\!\int_{0}^\infty\!\!\!\!\int_{0}^{\infty }\overline{v\left( t\right) }%
a\left( \tau \right) v\left( t-\tau \right)\,\D t\,\D\tau \,\geq\, 0
\quad \text{for all } v(t) \text{ with compact support.}
\end{equation}

A natural question to ask is whether every system of the form \eqref{open} whose friction function $a_1(t)$ satisfies the condition \eqref{nogain} is a subsystem of a conservative system.  The answer is positive, and there exists in fact a unique minimal extension up to isomorphism \cite{FigotinSchenker}.  This extension, or, equivalently, the form \eqref{delayedresponse}, is canonically constructible through the Fourier-Laplace transform $\hat a_1(\zeta)$ of $a_1(t)$.  It follows that all open systems of this type can be studied as a subsystem of a larger closed one.

This minimal conservative extension should be viewed as the space $H_1$ of observable variables coupled to the subspace of the original space of hidden variables $H_2$ that is detectable by the observable system; we denote this coupled subspace by $H_{2c}$. The influence on $H_1$ of this subsystem of hidden variables is manifest by $a_1(t)$ and reconstructible by $a_1(t)$, up to isomorphism.  The decoupled part of $H_2$, denoted by $H_{2d} = H_2 \ominus H_{2c}$, is not detectable by the reduced open system \eqref{reduced} in $H_1$.

The detectable part of the hidden variables may be a very restricted subspace of the naturally given space of hidden variables.  We will show that, if the coupling is of finite rank, in particular, if the observable system is finite dimensional, then the spectral multiplicity of the conservative extension is finite.   This leads to the following observation: Suppose our system of hidden variables is modeled by nearest-neighbor interactions in an infinite multidimensional lattice or the Laplace operator in continuous space, both of which have infinite multiplicity, and suppose that our observable system is a finite-dimensional resonator (perhaps very large, but finite).  Then there is a huge subspace of the hidden variables that is not detected by the resonator, in other words, there are many hidden degrees of freedom that are not detected by the resonator, and which, in turn, do not influence its dynamics.

\medskip

In this discussion, the roles of $H_1$ and $H_2$ may just as well be switched.  One may solve for $v_2(t)$ and obtain an analogous expression to \eqref{open} with delayed response function $a_2(t) = \Gamma^\dagger \E^{-\I\Omega_1 t} \Gamma$.  $H_1$ is then decomposed into its coupled and decoupled parts: $H_1 = H_{1c} \oplus H_{1d}$.

With respect to the decomposition of $\mathcal H$ into the coupled and decoupled parts of the observable and hidden variables,
\begin{equation}\label{decomposition}
\mathcal{H} = H_{1d}\oplus H_{1c}\oplus H_{2c}\oplus H_{2d},
\end{equation}
the frequency operator $\Omega$ for the closed system in $\mathcal H$ has the matrix form
\begin{equation}\label{OmegaDecomposition}
\Omega = \left[
\begin{array}{cccc}
\Omega_{1d} & 0 & 0 & 0 \\
0 & \Omega_{1c} & \Gamma_c & 0 \\
0 & \Gamma_c^\dagger & \Omega_{2c} & 0 \\
0 & 0 & 0 & \Omega_{2d} \\
\end{array}
\right].
\end{equation}
The minimal conservative extension of the system \eqref{reduced} in $H_1$ within the given system $(\mathcal{H},\Omega)$ is the space generated by $H_1$ through $\Omega$, or the \emph{orbit} of $H_1$ under $\Omega$, denoted by $\orbit{\Omega}(H_1)$.  Similar reasoning can be applied to $H_2$.   We therefore obtain
\begin{eqnarray}
\orbit{\Omega}(H_1) &=& H_1 \oplus H_{2c}, \\
\orbit{\Omega}(H_2) &=& H_{1c} \oplus H_2.
\end{eqnarray}
The orbit of a subset $S$ of $\mathcal H$ is
\[
\orbit{\Omega}(S) = \text{closure of } \left\{ f(\Omega) v \,|\, f\in C^\infty_0(\mathbb{R}), v\in S \right\}.
\]
If $\Omega$ is bounded, $\orbit{\Omega}(S)$ is equal to the smallest subspace of $\mathcal H$ containing $S$ that is invariant, or closed, under $\Omega$.  Equivalently, it is the smallest subspace of $\mathcal H$ containing $S$ that is invariant under $(\Omega-\I)^{-1}$; this latter formulation is also valid for unbounded operators.  The relevant theory can be found, for example, in \cite{AkhiezerGlazman} or \cite{ReedSimon}.

The closed subsystem $(H_{1c}\oplus H_{2c},\Omega_c)$ with frequency operator
\[
\Omega_c =
\left[
\begin{array}{cc}
\Omega_{1c} & \Gamma_c \\
\Gamma_c^\dagger & \Omega_{2c} \\
\end{array}
\right],
\]
is in fact \emph{reconstructible} by either of the open subsystems $(H_{1c},\Omega_{1c},a_1(t))$ or $(H_{2c},\Omega_{2c},a_2(t))$.  Equivalently, $(H_{1c}\oplus H_{2c},\Omega_c)$ is the unique minimal conservative
extension, realized as a subsystem of $(\mathcal H,\Omega)$, of each of its open components separately.  This motivates the following definition.

\begin{definition}[reconstructibility]
We call a system $(\mathcal{H},\Omega)$ together with the decomposition $\mathcal{H} = H_1\oplus H_2$ \emph{reconstructible} if $H_{1d}=0$ and $H_{2d}=0$, that is, $(\mathcal{H},\Omega)$ is the minimal conservative extension of each of its parts.
\end{definition}

The next theorem asserts the existence of a unique reconstructible subsystem of $(\mathcal{H},\Omega)$ that contains the images of $\Gamma$ and $\Gamma^\dagger$ and gives a bound on the multiplicity of $\Omega$, as we have discussed above. 
Define
\begin{equation}
\mathring\Omega =  \left[  
\begin{array}{cc}
\Omega_1 & 0 \\ 
0 & \Omega_2%
\end{array}
\right], \quad \mathring\Gamma =  \left[  
\begin{array}{cc}
0 & \Gamma \\ 
\Gamma^\dagger & 0%
\end{array}
\right]. 
\end{equation}

\begin{theorem}[system reconstruction]\label{theoremMultiplicity}
Define
\begin{align}
H_{2c} &= \orbit{\Omega}(H_1) \ominus H_1,
    & H_{2d} &= H_2 \ominus H_{2c}, \\
H_{1c} &= \orbit{\Omega}(H_2) \ominus H_2,
    & H_{1d} &= H_1 \ominus H_{1c}.
\end{align}
Then
\begin{equation}\label{one}
H_{1c}\oplus H_{2c} = {\mathcal{O}}_{\Omega}(H_{1c}) = {\mathcal{O}}_{\Omega}(H_{2c}) = {\mathcal{O}}_{\Omega}(\mathrm{Ran}\,{\mathring\Gamma}). 
\end{equation}
In particular, $H_{1c}\oplus H_{2c}$ is reconstructible and
\begin{equation}\label{two}
\mathrm{multiplicity}\,(\Omega_c) \,\leq\, \min\big(2\,\mathrm{rank}\,(\Gamma),\mathrm{dim}(H_{1c}),\mathrm{dim}(H_{2c})\big), 
\end{equation}
in which $\Omega_c$ denotes the restriction of $\Omega$ to $H_{1c}\oplus H_{2c}$.
\end{theorem}

\begin{proof}
That $H_{1c}\oplus H_{2c}$ is invariant under $(\Omega-\I)^{-1}$ (or $\Omega$, if $\Omega$ is bounded) and contains the range of $\mathring\Gamma$ is evident from the decomposition \eqref{OmegaDecomposition} of $\Omega$.
To prove the first equality in \eqref{one}, let
\[
\orbit{\Omega}(H_{1c}) = H_{1c}\oplus H'_{2c}, 
\]
in which $H_{2c} = H'_{2c} \oplus H''_{2c}$.  We see that
$
H_1 \oplus H'_{2c}
$
is closed under $(\Omega-\I)^{-1}$, and since $H_1 \oplus H_{2c}$ is the smallest subspace of $\mathcal H$ that is closed under $(\Omega-\I)^{-1}$, we have $H''_{2c}=0$.  The second equality in \eqref{one} is proved similarly.

Since $\range\mathring\Gamma\subseteq H_{1c}\oplus H_{2c}$, we have $\orbit{\Omega}(\range\mathring\Gamma) \subseteq H_{1c} \oplus H_{2c}$.
It remains to be proved that $H_{1c} \oplus H_{2c} \subseteq \orbit{\Omega}(\range\mathring\Gamma)$.  First,

\begin{eqnarray}
{\mathcal{O}}_{\Omega }(\mathrm{Ran}\,\mathring{\Gamma})
&=& {\mathcal{O}}%
_{\Omega ,\mathring{\Gamma}}(\mathrm{Ran}\,\mathring{\Gamma}) \\
 & & [\text{because }  \mathring{\Gamma}({\mathcal{O}}_{\Omega }(\mathrm{Ran}\,\mathring{\Gamma}))=
\mathrm{Ran}\,\mathring{\Gamma}\subseteq {\mathcal{O}}_{\Omega }(\mathrm{Ran}%
\,\mathring{\Gamma}) ] \notag \\
&=& {\mathcal{O}}_{\mathring{\Omega},\mathring{\Gamma}}(%
\mathrm{Ran}\,\mathring{\Gamma})\quad [\text{because }
\mathring{\Omega}=\Omega -\mathring{\Gamma}] \\
&=& {\mathcal{O}}_{%
\mathring{\Omega}}(\mathrm{Ran}\,\mathring{\Gamma}) \\
&=& {\mathcal{O}}_{\mathring{\Omega}}(\mathrm{Ran}%
\,\Gamma )\oplus {\mathcal{O}}_{\mathring{\Omega}}(\mathrm{Ran}\,\Gamma
^{\dagger })
\\ & &
[\text{because } \mathrm{Ran}\,\mathring{\Gamma}=\mathrm{Ran}\,\Gamma \oplus \mathrm{Ran}\,\Gamma ^{\dagger }] \notag \\
&=& {\mathcal{O}}_{\Omega _{1}}(\mathrm{Ran}\,\Gamma )\oplus {\mathcal{O}}%
_{\Omega _{2}}(\mathrm{Ran}\,\Gamma ^{\dagger }).
\end{eqnarray}
From this we see that $\orbit{\Omega_2}(\range\Gamma^\dag) \subseteq H_{2c}$.  Since
$\left(H_2\ominus \orbit{\Omega_2}(\range\Gamma^\dag)\right) \perp \range(\Gamma^\dag)$, we have that $H_1 \oplus \orbit{\Omega_2}(\range\Gamma^\dag)$ is $(\Omega-i)^{-1}$-invariant, so that $H_{2c} \subseteq \orbit{\Omega_2}(\range\Gamma^\dag)$ by the minimality of $H_1\oplus H_{2c}$ with respect to closure under $(\Omega-i)^{-1}$. Therefore, $H_{2c} = \orbit{\Omega_2}(\range\Gamma^\dag)$; similarly, $H_{1c} = \orbit{\Omega_1}(\range\Gamma)$.  We conclude that  $H_{1c}\oplus H_{2c}  = {\mathcal{O}}_{\Omega}(\mathrm{Ran}\,{\mathring\Gamma})$, and this finishes the proof of \eqref{one}.

To prove \eqref{two}, note that the multiplicity of $\Omega_c$ is the minimal number (which could be infinity) of generating vectors needed to generate $H_{1c}\oplus H_{2c}$ by $\Omega_c$, or, equivalently, by $\Omega$.  Thus, by \eqref{one}, the multiplicity of $\Omega_c$ is bounded by the dimension of $H_{1c}$, the dimension of $H_{2c}$, and the dimension of the range of $\mathring\Gamma$.  Since the dimensions of the ranges of $\Gamma$ and $\Gamma^\dagger$ are equal and
$\mathrm{Ran}\,\mathring{\Gamma}=\mathrm{Ran}\,\Gamma \oplus \mathrm{Ran%
}\,\Gamma ^{\dagger }$, we see that the range of $\mathring\Gamma$ has twice the dimension of the range of $\Gamma$.  This completes the proof of the Theorem.

\end{proof}

\begin{example}[lattice]\label{exampleLattice}
Let $\mathcal H$ be the Hilbert space of square-summable complex-valued functions on the integer lattice $\mathbb{Z}^3 = \{n=(n_1,n_2,n_3) \,|\, n_1,n_2,n_3\in\mathbb{Z}\}$,
\[
\mathcal{H} = \left\{ f:\mathbb{Z}^3\to\mathbb{C}\, |\, \sum_{n\in\mathbb{Z}} |f(n)|^2 <\infty \right\},
\]
and let $\Omega$ be the discrete Laplace operator:
\[
\Omega f = \sum_{j=1}^3 \Delta_j f,
\]
in which $(\Delta_j f)(n) = f(n+e_j) - 2f(n) + f(n-e_j)$ and $e_j$ is the $j$-th elementary vector in $\mathbb{Z}^3$ ({\it e.g.,} $e_1 = (1,0,0)$).

Let $H_1$ be the finite-dimensional subspace of $\mathcal H$ consisting of complex-valued functions on the lattice cube
\[
\mathcal{Q} = \left\{ n=(n_1,n_2,n_3) \,|\, 0\leq n_j <N, j=1,2,3 \right\},
\]
which is isomorphic to $\mathbb{C}^{N^3}$.  Since $\Omega$ involves only nearest-neighbor interactions, the range of $\Gamma$ is the space of complex-valued functions on the surface of $\mathcal Q$, which has dimension $6N^2 -12N +8$.  Therefore, by Theorem \ref{theoremMultiplicity}, the multiplicity of the restriction of $\Omega$ to the minimal conservative extension of $H_1$ in $\mathcal H$ is no greater than
$12N^2 - 24N + 16$.  However, the multiplicity of $\Omega$ in $\mathcal H$ is infinite, showing that the restriction of $\Omega$ to $H_{2d}$, the decoupled part of $H_2 = \mathcal{H}\ominus H_1$ has infinite multiplicity.  This is a very large space of degrees of freedom that are not detected by the cube $\mathcal Q$ and therefore do not influence its dynamics.

\end{example}

\section{Discussion}

Based on the minimal conservative extension of an open system, we develop a clear mathematical framework for the widely used concept of coupling.  In this paper, we have focused on the amount of information about a conservative system that is encoded in a given open subsystem and the reconstruction of that part of the conservative system that is equivalent to the abstract minimal extension.  The efficiency of the construction is demonstrated by a concrete statement showing, by analysis of spectral multiplicity, that very often this extension is a very small part of the conservative system.  In ongoing work, we analyze the interaction between the spectral theories of the internal dynamics of two systems and a coupling operator between them and its bearing on the decomposition of an open system into dynamically independenty parts.

To give a sense of the potential of the approach, we mention as problems that are naturally addressed in the framework of conservative extensions
(1) the classification and analysis of eigenmodes and resonances,
(2) applications to the construction of dynamical models for thermodynamics, and
(3) transmission of excitations in complex inhomogeneous media.

An interesting conclusion of our studies of coupling of open systems and the
spectral multiplicity is that there can be degrees of freedom which are
completely decoupled from the rest of the system. Since the spectral
multiplicity is a consequence of a system's natural symmetries, one can
consider such a decoupling as an explanation for so-called ``frozen" degrees
of freedom observed in the treatment of the specific heat for crystalline
solids (Dulong-Petit law) \cite[Section 3.1]{Gallavotti}. The analysis of the
specific heat involves the law of equipartition of energy and the number of
degrees of freedom, and in order to agree with the experiment one has to
leave out some degrees as if they were not excited and can be ``frozen", 
\cite[Section 3.1]{Gallavotti}, \cite[Section 6.4]{Huang}.



\begin{thebibliography}{}
%
%

\bibitem{AkhiezerGlazman} Akhiezer, N. I., Glazman, I. M.:
Theory of Linear Operators in Hilbert Space.
Dover, New York (1993)

\bibitem{FigotinSchenker} Figotin, A., Schenker J.:
Spectral Theory of Time Dispersive and Dissipative Systems.
J. Stat. Phys.
\textbf{118} (1), 199--263 (2005)

\bibitem{Gallavotti} Gallavotti, G.:
 Statistical Mechanics, A Short Treatise.
Springer, Berlin (1999)

\bibitem{Huang} Huang, K.:
Statistical Mechanics.
Wiley (1987)

\bibitem{ReedSimon} Reed, M., Simon, B.:
Functional Analysis, Vol. I.
Academic Press, New York (1972)

\bibitem{ShipmanVenakides} Shipman, S.P., Venakides, S.:
Resonant transmission near nonrobust periodic slab modes.
Phys. Rev. E., \textbf{71}, 026611-1--10 (2005)

\bibitem{Tikhodeev}
Tikhodeev, S.G., Yablonskii, A.L., Muljarov, E.A., Gippius, N.A., Ishihara, T.:
Quasiguided modes and optical properties of photonic crystals slabs.
Phys. Rev. B, \textbf{66}, 045102 (2002)

\bibitem{Tip}  Tip, A.:  Linear absorptive dielectrics.  Phys. Rev. A, \textbf{57}, 4818--4841 (1998)

\end{thebibliography}
\end{document}